\documentclass[msmath,amssymb,aip,rsi,twocolumn,epsfig,showpacs,bibliography,lengthcheck,superscriptaddress]{revtex4-2}

\usepackage{graphicx}
\usepackage{dcolumn}
\usepackage{bm}

\usepackage{graphicx}
\usepackage{dcolumn}
\usepackage{bm}
\usepackage{hyperref}
\usepackage{epstopdf}
\usepackage{subcaption}
\usepackage{caption}
\usepackage{footmisc}
\usepackage{amsmath}
\usepackage{soul}
\usepackage{color}
\usepackage{appendix}
\usepackage{natbib}

\newcommand{\superrefcite}[1]{%
	\textsuperscript{Ref. \unskip\csname citenum\endcsname{#1}}%
}

\begin{document}

\title{Applications of silicon carbide {as window} materials in atomic cells and atomic devices}
\newcommand{\arccot}{\mathrm{arccot}\,}

\author{Z.-P. Xie}
\affiliation{Department of Precision Machinery and Precision Instrumentation, Key Laboratory of Precision Scientific Instrumentation of Anhui Higher Education Institutes, University of Science and Technology of China, Hefei 230027, China}
\author{C.-P. Hao}
\email{cphao@mail.ustc.edu.cn}
\affiliation{Deep Space Exploration Laboratory, Hefei 230088, China}

\author{D. Sheng}
\email{dsheng@ustc.edu.cn}
\affiliation{Department of Precision Machinery and Precision Instrumentation, Key Laboratory of Precision Scientific Instrumentation of Anhui Higher Education Institutes, University of Science and Technology of China, Hefei 230027, China}
\affiliation{Hefei National Laboratory, University of Science and Technology of China, Hefei 230088, China}

\begin{abstract}

Atomic cells made by anodically bonding silicon and borosilicate glasses are widely used in atomic devices. One inherent problem in these cells is that the silicon material blocks beams with wavelengths shorter than 1000 nm, which limits available optical accesses when alkali metal atoms are involved. In this work, we investigate the possibility of the silicon carbide material as an alternative of silicon materials in fabricating anodically bonded cells. We demonstrate that the optical, thermal and mechanical properties of silicon carbide help to improve the performance of atomic devices in certain applications.
\end{abstract}

\maketitle

\section{Introduction}
Atomic devices are miniaturized instruments based on atomic spectroscopy~\cite{kitching2018}, which have wide applications in time keeping~\cite{knappe2004,newman2019}, and sensing electric fields~\cite{jing2020,sedlacek2012}, magnetic fields~\cite{shah2007,budker2013}, rotations~\cite{walker16}, and accelerations~\cite{biedermann2017}. To achieve high performances of atomic devices, high-quality atomic vapor cells often play critical roles. Currently, the most widely used method to make miniaturized or standardized atomic vapor cells is connecting borosilicate glasses with silicon wafers using the anodic bonding method, which was first developed in 1960s~\cite{wallis1969} and then introduced to atomic devices in 2004~\cite{liew2004}. Some other novel methods were also developed in the past two decades for massive production of miniaturized atomic vapor cells, such as the micro-glass-blowing method~\cite{eklund2008} and the additive manufacturing based approach~\cite{wang2024}.

While the glass-silicon bonding techniques have achieved great success in atomic cells, there are still some drawbacks. A common problem is that the silicon material has a band gap of 1.2 eV, and blocks light beams with wavelengths shorter than 1000 nm. This problem limits the available optical windows for measurements when alkali metal atoms are involved. A possible solution is brought up by Romalis' group~\cite{nezih2014} to use one of the second-generation semiconductor materials, {gallium phosphide} (GaP), as an alternative to silicon for fabrications of atomic vapor cells. In comparison with Si, GaP has a larger band gap (2.2 eV), and a similar thermal expansion coefficient.

In this paper, we consider a third-generation semiconductor material, silicon carbide (SiC), as a replacement of Si in anodically bonded cells. SiC materials are often viewed as important options for uses in extreme conditions. For example, SiC shows a relatively low helium permeability at temperatures up to {900} $^\circ$C~\cite{hino2005}, and has better radiation resistance compared to traditional glasses~\cite{owens2004}. SiC also has good mechanical properties, including a strong strength to resist deformation. This makes it a good candidate to make ultra-thin windows of atomic cells which are required in some fundamental physics research work~\cite{shortino2024}. In these applications, the deformation of thin windows is a serious issue to affect the measurement precision due to the pressure difference between the inside and outside of the cell. Following this introduction, Sec. II describes the SiC-window cells made by anodic bonding technique, Sec. III covers the preliminary applications of these cells based on the thermal and optical properties of SiC, and Sec. V outlooks future work in related field.

\section{Anodically Bonded SiC-window cells}
{Table~\ref{tab:prop} shows some important physical properties of different materials~\cite{nezih2014,choyke1969,shaffer1971,levinshtein2001,neumeier2024,zanatta2019,glassbrenner1964,okada1984,pxprop}, including SiC used as a window material in atomic cells}. SiC has a band gap width of 3.26 eV~\cite{choyke1969}, so that it is transparent to beams with even shorter wavelengths ($\lambda >$ 380 nm) compared with GaP ($\lambda >$ 550 nm). In this paper, we focus on the widely-used hexagonal configuration of SiC (4H-SiC), whose index of refraction is about 2.6 for near infrared light beams. { Owing to its lower refractive index, raw SiC provides roughly 20\% superior optical transmission to raw GaP at the Rb D1 transition wavelength ($\lambda$=795 nm).} The beam transmission through the SiC window can be improved to 98.8\% when anti-reflection coatings {composed of tantalum and silicon dioxide} are added to both sides of the window, and this transmission value remains the same even we add a layer of Al$_2$O$_3$ in the outermost layer of the coating to prevent chemical reactions with alkali metals at high temperature~\cite{nezih2014}. 

\begin{table*}
	\caption{\label{tab:prop}Comparison of some important properties among different materials.}
	\begin{ruledtabular}
		\begin{tabular}{cccccccc}
			&band gap&index of refraction&thermal conductivity &thermal expansion&{He permeability}\\
			&(eV)&@ 800 nm &(W/(m $\cdot$ K))&coefficient ($10^{-6}$/K)&{(cm$^2$/s)} \\
			\hline
				SiC&3.26~{\superrefcite{choyke1969}} &2.60~{\superrefcite{shaffer1971}} &490~{\superrefcite{levinshtein2001}}&2.4~{\superrefcite{neumeier2024}}& {$<$10$^{-8}$ @ 900~$^\circ$C~\superrefcite{hino2005}}\\
			GaP&2.20~\superrefcite{nezih2014}&3.20~\superrefcite{nezih2014} & 110~\superrefcite{nezih2014} &4.8~\superrefcite{nezih2014}&
{$(4 \pm 4)\times 10^{-11}$ @ 200~$^\circ$C~\superrefcite{nezih2014}}\\
			Si&1.12~{\superrefcite{zanatta2019}}&-& 130~\superrefcite{glassbrenner1964} &2.6~\superrefcite{okada1984}&
{$3.3 \times 10^{-12}$@ 1200~$^\circ$C~\superrefcite{WIERINGEN1956}}\\
&&& &&{$4.8 \times 10^{-17}$@ 500~$^\circ$C~\superrefcite{rushton2014,alatalo1992}}\\
			Pyrex glass &-&1.47~\superrefcite{pxprop}& 1.1~\superrefcite{pxprop} &3.3~\superrefcite{pxprop}&
{$4.5 \times 10^{-9}$@ 200~$^\circ$C~\superrefcite{altemose1961}}\\					
		\end{tabular}
	\end{ruledtabular}
\end{table*}

\begin{figure}
	\includegraphics[width=3in]{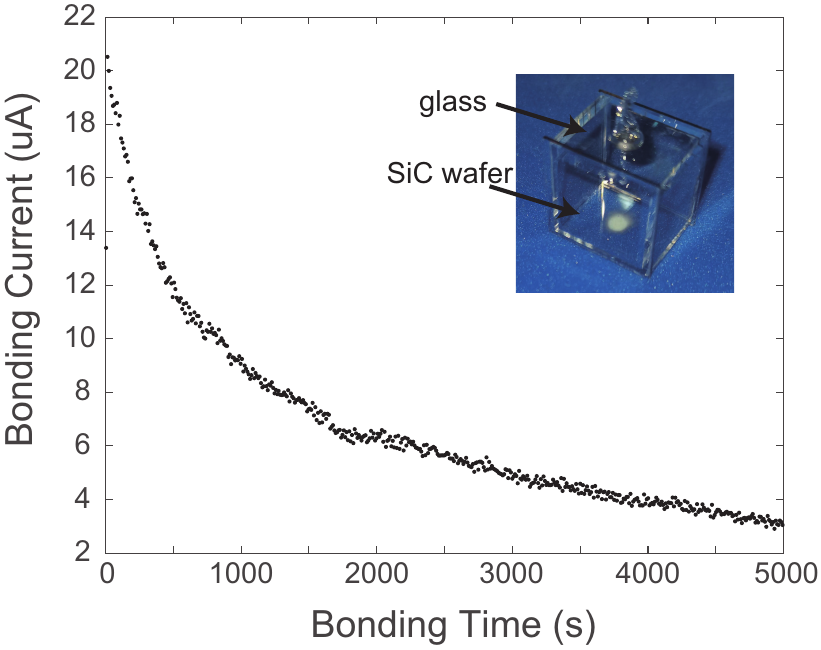}
	\caption{\label{fig:ad} Relation between of the bonding current and time during the process of anodically bonding a SiC window with Pyrex glasses.}
\end{figure}

Besides the good optical properties, the thermal expansion coefficient of SiC is also close to that of borosilicate glass according to Tab.~\ref{tab:prop}, therefore SiC windows can be anodically bonded with borosilicate glasses similar to Si and GaP. In the experiment, we bond a  0.5-mm-thick {anti-reflection coated} SiC window with Pyrex glasses under an electric field strength of about 8 kV/cm (with the voltage between electrodes as 1200 V and the distance between electrodes as 1.5 mm) and temperature of 300 $^\circ$C.  During the bonding process, ions migrate in the vicinity of the bonding surface, which is manifested macroscopically as a change in the bonding current. Figure~\ref{fig:ad} shows the monitored bonding current as a function of time during the bonding process. The magnitude of the bonding current is similar to that in the Si-glass bonding process under the same experimental conditions, and the anodic bonding process is terminated after the bonding current drops to a level below 10\% of its initial value.

\begin{figure}
	\includegraphics[width=3in]{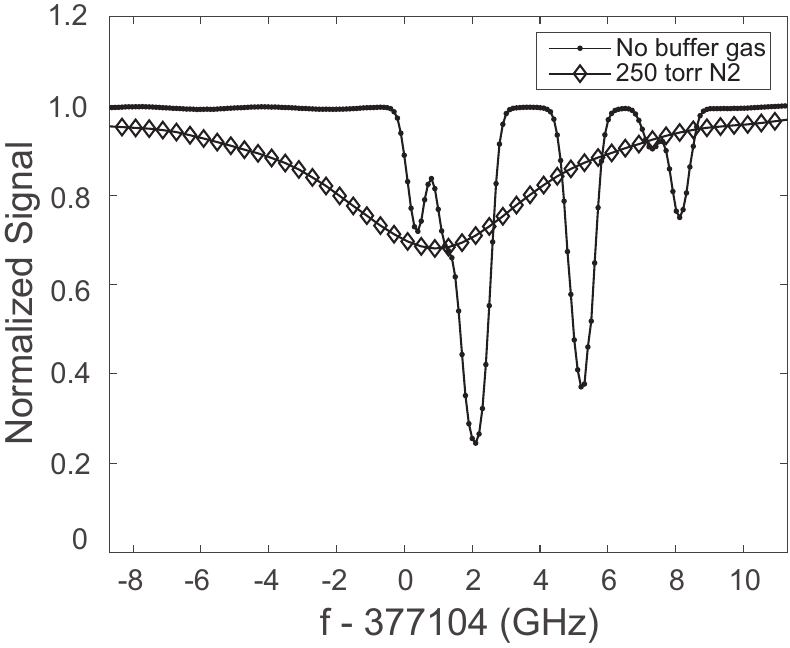}
	\caption{\label{fig:absorption} Comparisons of light absorption spectra near Rb D1 line using SiC-window based vapor cells with and without buffer gases. All data sets are acquired with a cell temperature of 80 $^\circ$C, and a beam with a diameter of 2 mm and a power of 0.2 mW.}
\end{figure}

One example of the bonded cell is shown in the inset of Fig.~\ref{fig:ad}, where two {anti-reflection coated} SiC windows are bonded on two opposite sides of the atomic cell. The external dimension of the cell is 16 mm $\times$ 16 mm $\times$ 13 mm, and the size remains the same for the bonded cell in the rest of the paper. To test the basic performance of these atomic cells with SiC windows, we have prepared two cells filled with different components: one is filled with only Rb atoms in natural abundance, and the other one contains Rb atoms and 250 torr N$_{2}$. Figure~\ref{fig:absorption} shows the optical absorption spectra of both atomic vapor cells by sending the light through the bonded SiC windows.  The vapor cell without buffer gases shows D1 transitions from the ground hyperfine states of both Rb isotopes, and the transition line width is limited by Doppler effects. For the cell filled with buffer gases, the D1 transition line width (full width of half maximum) is pressure broadened to 6.5 GHz, so that none of the ground hyperfine states can be resolved. In addition, its center frequency is red shifted by 2.5 GHz due to the collisions with buffer gas atoms. Both the pressure broadened line width and the collision shift agree with the results in Ref.~\cite{romalis1997}.

\section{Preliminary applications in atomic devices}
\begin{figure}
	\includegraphics[width=3in]{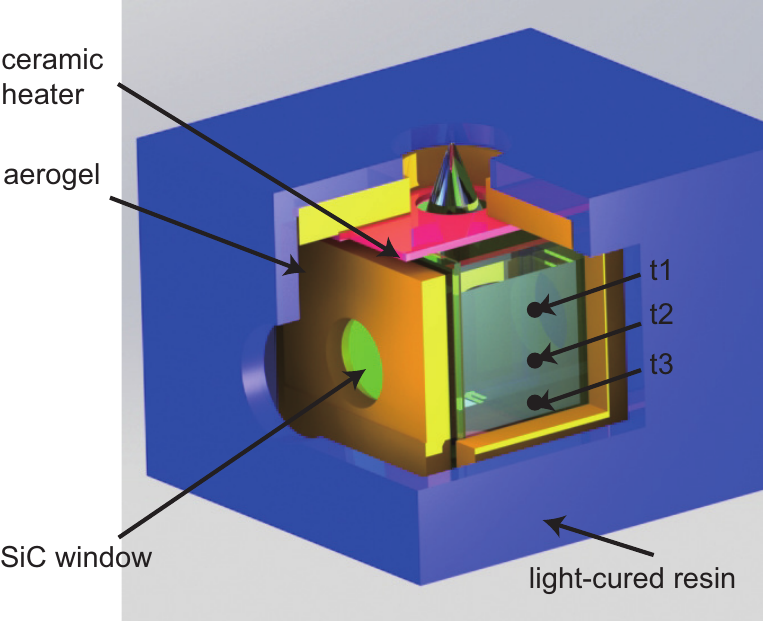}
	\caption{\label{fig:thermal} Illustration of the SiC-window based vapor cell used in the temperature gradient test, with the presence of heaters {(the bottom heater is not shown)}, thermal insulation materials and a container. The thermal couples are placed on the side window made of Pyrex glasses.}
\end{figure}

For compact atomic devices, atomic cells are often heated by attaching heaters directly to cell surfaces. Due to the non-uniform distribution of the heat source, this operation can lead to non-negligible temperature gradient inside the cell. Such a temperature gradient not only causes an inaccuracy of temperature reading, but also results in redistribution of alkali atoms, which would modify the long-term performance of the device or block the light during the atom migration process.  One way to reduce this temperature gradient is to improve the thermal conductivity of the cell material. According to the data in Tab.~\ref{tab:prop}, the thermal conductivity of SiC is as high as 490 W/(m $\cdot$ K) at room temperature, which is 3.3 times higher than that of Si, 4.5 times higher than that of GaP, and more than 400 times higher than that of Pyrex glasses.  Experiment tests are performed by comparing temperature gradients in two kinds of cells: one is the anodically bonded SiC-window based cell, and the other one is a pure Pyrex glass cell with an external dimension of 15 mm $\times$ 15 mm $\times$ 15 mm. {Both cells are filled with the same pressure (250 torr) of N$_2$.} As shown in Fig.~\ref{fig:thermal}, two ceramic heaters are placed at the top and bottom sides of each cell. Each cell is placed inside a 3D printed container made of light-cured resins, and the heat insulation materials in between the cell and the container are made of aerogel, which has a thermal conductivity around 0.025 W/(m $\cdot$ K). {Three T-type thermocouples, labeled $t_1$, $t_2$, and $t_3$, were vertically aligned on one Pyrex-glass window of the vapor cell to measure the temperature gradient. The distances between $t_1$–$t_2$ and $t_2$–$t_3$ were 4.0 mm and 4.7 mm, respectively, with an estimated uncertainty of 1 mm. With the cell temperature stabilized at 90 $^\circ$C using the reading from $t_2$, the temperatures at $t_1$ and $t_3$ were measured. For the SiC-window cell, these were 90.6 ± 0.2 $^\circ$C and 91.8 ± 0.2 $^\circ$C, respectively. In contrast, the Pyrex glass cell exhibited a much larger thermal gradient, with temperatures of 104.8 ± 0.2 $^\circ$C at $t_1$ and 82.1 ± 0.2 $^\circ$C at $t_3$. {The large thermal gradient observed on the Pyrex window arises from natural convection within the gas-filled cell. The Pyrex glass, due to its low thermal conductivity, is unable to conduct heat effectively enough to compensate for this gradient. For the SiC window, its high thermal conductivity dominates over convective effects, leading to a nearly uniform temperature distribution, and the minor remaining temperature difference between top and bottom parts is attributed to slight variations in heater contact.}

Finite-element simulations are performed to obtain the theoretical temperature distribution. The simulated temperatures for the SiC-window cell at the $t_1$ and $t_3$ are 91.0 $^\circ$C and 89.6 $^\circ$C, while those for the Pyrex-type cell are 97.5 $^\circ$C and 86.1 $^\circ$C. Both the experimental and simulation results confirm that the use of SiC material reduces the temperature gradient across the vapor cell by one order of magnitude compared to the conventional Pyrex design. The observed discrepancies between the experimental and simulation results are likely due to the simplified assumption of identical heating power and heat dissipation conditions for both ceramic heaters in the simulation, which does not fully capture the complexity of the actual physical system.}

\begin{figure}
	\includegraphics[width=3in]{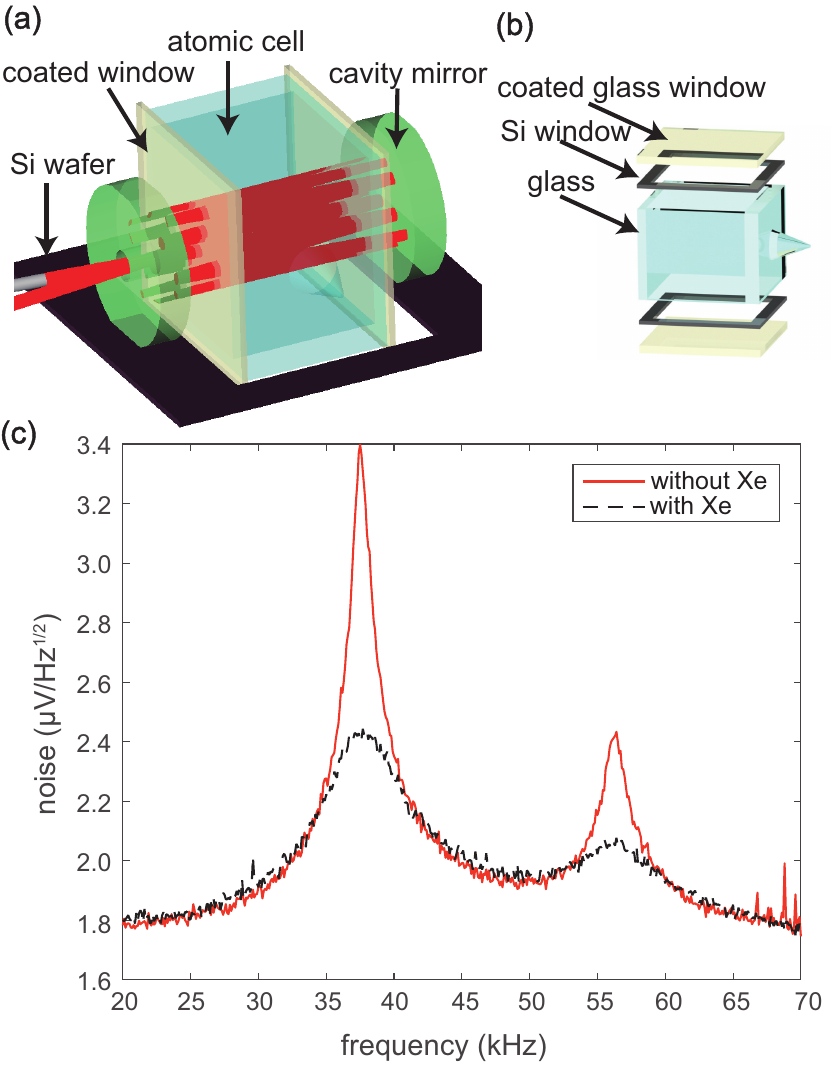}
	\caption{\label{fig:mp} (a) Illustration of the core optical setup in the new generation of Herriott-cavity-assisted nuclear spin comagnetometers. (b) Atomic cells made by bonding {anti-reflection coated} glass windows to the glass cell via Si wafers. (c) Spin-noise spectrum of Rb atoms using the setup in plot (a), with a cell temperature of 95 $^\circ$C, a probe beam power of 1 mW exiting from the cavity, and a probe beam detuning of -80 GHz from the Rb D1 line. Two cells are used: one is the buffer-gas cell without Xe gases used in Fig.~\ref{fig:absorption}, and the other one is the comagnetometer cell filled with Xe gases mentioned in the main text.}
\end{figure}

Another important class of atomic devices is the nuclear spin comagnetometer, which has wide applications in atomic gyroscopes~\cite{walker16} and fundamental physics researches~\cite{terrano2021}. Currently, the most commonly used nuclear spin comagnetometer is a hybrid system consisting of Xe isotopes and alkali-metal atoms. Here, the polarized alkali-metal atoms are used to both hyperpolarize and probe the nuclear spins. A key problem in this system is the sensitivity of the in-situ alkali-metal-atom magnetometer is limited by collisions with Xe atoms. Recently, multipass cells~\cite{Silver2005} are introduced to the nuclear spin comagnetomters to improve the light-atom interaction length~\cite{hao2021}. Compared with the commonly used parametric modulation method~\cite{walker16}, this passive method increases the detection sensitivity without introducing any extra instability factors to the comagneotmeter system~\cite{gao2024}. The multipass cavity is originally placed inside the atomic cell for easy operations, but this configuration has several disadvantages. As the size of the cell is reduced to make the system more compact, collisions between the nuclear spin atoms and the cavity mirrors become an important source for nuclear spin depolarization. Moreover, the minimum size of the cell is limited by the cavity length. A new generation of multipass-cavity-assisted comagnetometers is designed to avoid these problems by placing the cavity mirrors outside the cell as shown in Fig.~\ref{fig:mp}(a), where cavity mirrors are bonded on a piece of silicon wafer so that the cavity as an integrated element can be conveniently combined with a 3D printed platform for use.

We find that SiC windows are particulary useful for atomic cells in the new generation of multipass-cavity-assisted comagnetometers. As mentioned above, this upgraded system requires the cell windows where cavity beams transmit through to be anti-reflection coated on both sides. Using the traditional Si-material based anodic bonding technique, this can be achieved by connecting {anti-reflection coated} glass windows and a glass cell with silicon wafers in the middle, as shown in Fig.~\ref{fig:mp}(b). While we have successfully implemented this scheme in practice, we also find its disadvantage that the fabrication procedure involves a multi-step bonding process, which increases the fabrication risks and reduces the efficiency.  Instead, the SiC material helps to simplify the cell fabrication processes, where one SiC window is attached with the glass cell by just one step of bonding as demonstrated in Fig.~\ref{fig:ad}.   On the other hand, the high transmission value of {anti-reflection coated} SiC windows also make it suitable for multipass applications. 

An SiC-window based atomic cell is prepared for the nuclear spin comagnetometer, where this cell has the same size as the previously prepared ones and is filled with 4 torr $^{129}$Xe, 30 torr $^{131}$Xe, 200 torr N$_{2}$, and Rb atoms in natural abundance. The {anti-reflection coated}-SiC-window based cell is placed the inside a multipass cavity as shown in Fig.~\ref{fig:mp}(a). An input light beam with a wavelength of 795 nm enters the cavity through a hole in the center of the front cavity mirror, reflects between the cavity mirrors 22 times and passes through both SiC windows with a total number of 44 times before exiting the cavity through the same hole in the front cavity mirror. The recorded transmitted beam power in presence of the SiC-window cells is about 50\% of the case without the cell, where the transmission of the SiC-window is reduced due to accumulation of Rb atoms on window surfaces. The spin-noise spectrum of Rb atoms in this comagnetometer cell is measured based on aforementioned setup, and its result is compared with the case using a cell without Xe atoms in Fig.~\ref{fig:mp}(c). The two peaks in the noise spectrum correspond to the atomic projection noises of two isotopes of Rb atoms~\cite{Li2011}. These comparisons demonstrate the advantage of multipass cells which help make the atomic noises visible, and confirm the broadening of Rb magnetometer line width with the presence of Xe atoms.

\begin{figure}
	\includegraphics[width=3in]{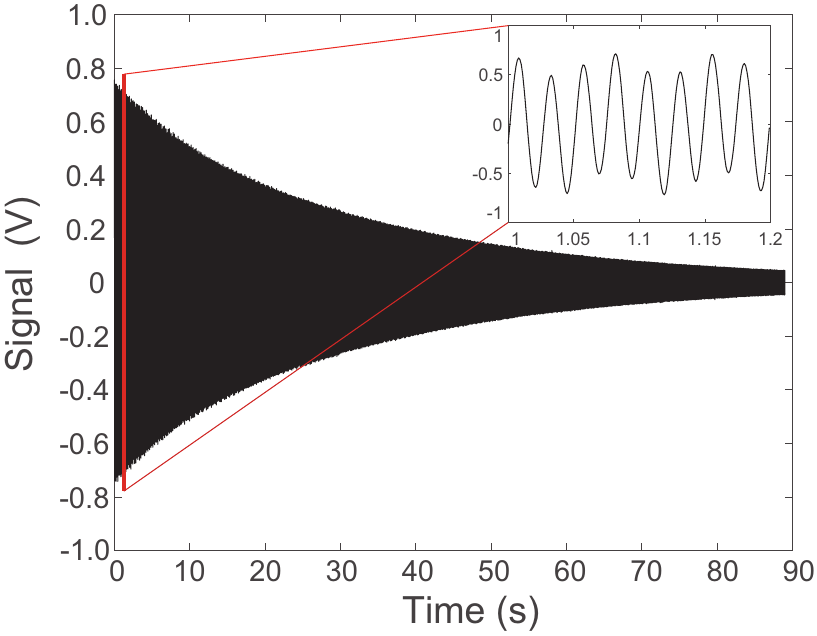}
	\caption{\label{fig:comag} Xenon isotope precession signals probed by the in-situ Rb magnetometer.}
\end{figure}

We also test the collision properties between the cell inner surfaces and Xe atoms, which are important factors to determine Xe depolarization time.  To accurately measure Xe depolarization time in this comagnetometer system, we adapt the pulsed pump-probe scheme, where the pumping beam is polarization modulated in the probe stage so that the effective fields from polarized Rb is largely time averaged out~\cite{feng2022}. Figure~\ref{fig:comag}(a) shows the experiment results of the Xe isotope signals probed by the in-situ Rb magnetometer, with a bias magnetic field of 35 mG and a cell temperature of 110 $^\circ$C. Since the vapor cell has a relatively large dimension and symmetric shape, the electric quadrupole splitting of $^{131}$Xe is not experimentally resolved~\cite{feng2020}. In this case, Xe signals can be well fitted by two damped oscillating functions, from which we extract transverse depolarization time of $^{129}$ Xe and $^{131}$Xe as 33 s and 13 s, respectively. These depolarization values for the SiC-window based cell agree well with the results using a Pyrex glass cell under the same experiment conditions.

\section{Conclusion}
In conclusion, we have demonstrated the feasibility of replacing Si by SiC in anodically bonded atomic cells, and these SiC-window based cells show good thermal and optical properties which are particularly useful for the multipass-cavity-assisted atomic comagnetometers. In the future, we are planning to implement the SiC-window based atomic cells in the new generation of nuclear spin comagnetometers for precision measurements~\cite{zhang2023}, where the multipass cavity is designed to be placed outside of the atomic cell.

\section*{acknowledgements}
This work was partially carried out at the USTC Center for Micro and Nanoscale Research and Fabrication. This work was supported by Natural Science Foundation of China (Grant No. 12174372).

\section*{Data Availability}
The data that support the findings of this study are available from the corresponding author upon reasonable request.

%

\end{document}